# Location as a service with a MEC architecture

Christopher Schahn, Jorin Kouril
Fraunhofer FOKUS
Berlin, Germany
firstname.lastname@fokus.fraunhofer.de

Bernd Schaeufele, Ilja Radusch
Daimler Center for Automotive IT Innovations
TU Berlin
Berlin, Germany
firstname.lastname@dcaiti.com

*Abstract*—In recent years, automated driving has become viable, and advanced driver assistance systems (ADAS) are now part of modern cars. These systems require highly precise positioning. In this paper, a cooperative approach to localization is presented. The GPS information from several road users is collected in a Mobile Edge Computing cloud, and the characteristics of GNSS positioning are used to provide lane-precise positioning for all participants by applying probabilistic filters and HD maps.

*Keywords—cooperative positioning, mobile edge computing, cooperative ITS*

## I. Introduction

A lot of effort has been put into the development of automated driving, both in industry and in research. Nowadays, level 3 automated driving [1] for specific operational design domains [2] is already commercially available by several car manufacturers, such as Mercedes Benz' Drive Pilot [3]. Yet, for fully automated driving, there are still challenges, especially in urban environment.

Additionally, Advanced Driver Assistance Systems (ADAS), e.g., lane keeping assistant, or adaptive cruise control, have become more and more common. Thereby, various sensors, such as camera or RADAR, have found their ways into many modern cars. Furthermore, many cars have Global Positioning System (GPS) sensors on board for navigation and other systems.

In the last decades, vehicle-to-vehicle and vehicle-to-infrastructure communication (V2X) has been an ongoing research field. It is also integrated in modern series-production vehicles, such as the Volkswagen Golf 8 [4]. V2X enables many cooperative driver assistance systems (CoDAS) [5], such as cooperative blind-spot assistant [6] or cooperative adaptive cruise control [7]. These systems benefit from the V2X's advantage of not requiring line-of-sight, unlike most other sensors.

With the evolution of automated driving and V2X, a new domain has been established: cooperative driving. Many challenging situations for automated vehicles that are difficult with sensors only can be resolved by cooperative maneuvers. There are different forms, from long-term cooperation, such as highway platooning [8], to short-term negotiated maneuvers, e.g., merging at highway access ramps [8].

There are several ways to organize cooperative driving maneuvers. One solution is to calculate trajectories for maneuvers and inform other vehicles about the intention. When the other vehicles agree and respond with matching trajectories, it is safe to drive the maneuver [9]. Other approaches are based on explicit negotiation and defined roles for the participating cars in a distributed state machine [10].

Technologies such as automated driving or V2X can be found in modern series production cars already, while cooperative driving is still in research. However, in projects with field tests, such as TEAM [11] or IMAGinE [12], the feasibility of this technology has been proven. The possibilities of cooperative driving for complex maneuvers allow a wide range of usage of this technology.

At the same time, car manufacturers equip their cars with cellular communication and provide additional assistance through custom backend services. The introduction of 5G communication also facilitates the rollout of cloud-based services. For more time critical applications, Mobile Edge Computing (MEC) provides a novel way of interaction between road users and services, so that location-based services can be supported by the infrastructure [8].

Automated driving, ADAS, and CoDAS rely on two essential technologies. For the planning of trajectories and for lane-precise assistance systems, a high definition (HD) map is required, which provides information about lanes and their relation to each other, e.g., in intersection areas. In order to make full usage of HD maps, it is necessary to have highly precise positioning available. The localization of a road user must be at least lane-level precise. One example for this is a cooperative traffic light assistant, which requires the system to know on which lane the road user is on.

Most modern vehicles have a Global Navigation Satellite System (GNSS) sensor available, e.g., GPS. Furthermore, other road users, such as cyclists or users of micro-mobility, often have a GPS receiver in form of a smartphone in usage during rides. However, GNSS sensors suffer from various problems that result in an inaccuracy of up to several meters [13]. There are three sources for GPS errors: satellite-based errors, signal propagation errors, and receiver errors. Satellite-based errors are mostly caused by inaccuracy of the internal satellite clocks. Atmospheric disturbances in the ionosphere and troposphere are the reason for the second category of errors. Receiver-based errors occur in the GPS device, e.g., by clock drifts or wrong calibration. The satellite and atmospheric error sources are similar to all nodes in a certain area. Therefore, knowledge about the position from several road users and additional information, such as HD maps, allow for localization improvement in cooperative systems.

In this paper with Location as a Service (LaaS), a novel MEC cloud-based approach to improve the localization of connected road users is presented. By using the anonymized positions of several users and projecting them onto a HD map, the most likely lane of each participating user can be estimated with a probabilistic filter. This highly accurate lane position can be used in different assistance systems in vehicles and on smart phones for other road users.

This paper is structured as follows: In the next section, an overview of the related work to our approach is given. In section 3, the architecture of the MEC cloud solution is presented. The MEC-based LaaS is then explained in detail in section 4. Following this, in section 5 the solution is evaluated and the results are discussed. An overview and an outlook on future developments are given in section 6.

The work presented in this paper was conducted in the KIS'M project, funded by the German Federal Ministry for Digital and Transport (BMDV).

## II. RELATED WORK

GNSS systems are prone to positioning errors. In general, there are three systematic errors: satellite-based, signal propagation and receiver-based errors. In [13], an approach is explained, in which vehicles exchange the measured pseudo-ranges to satellites, with which they are able to eliminate all common errors, i.e., all but the user-based errors. This method makes usage of the double difference by measuring the pseudo-range to a common set of satellites on a group of vehicles close to each other. As the satellite-based and signal propagation errors are the same for these vehicles, the double-difference calculation removes them so that a precise relative position between these vehicles can be calculated. This solution applies Dedciated Short Range Communication (DSRC), so only users with respective hardware and a direct communication channel are able to benefit.

A detailed evaluation of the double-difference approach is undertaken in [14]. In this paper, a weighted variation of the double-difference algorithm is presented, using the covariance of the GPS measurements. Moreover, the authors introduce an extended algorithm, where vehicles not only share the pseudo-ranges, but also the calculated relative vector between each other. By using several relative vectors and road space constraints, the true absolute position can be estimated as well. In the evaluation, the solution with DSRC is compared to an approach that is centralized, just as LaaS. It is shown that centralized approaches outperform the distributed ones, as they can make use of more available input data.

A hierarchical approach for relative cooperative positioning is presented in [15]. There are branch and leaf nodes, in which all leaf nodes are vehicles that are associated to a specific branch node vehicle. The leaf nodes with their respective branch node calculate the relative positioning between each other, making usage of common pseudo-range measurements. Moreover, a cooperative map-matching (CMM) algorithm is applied to estimate absolute positions and thereby a pseudo-range correction. The branch nodes exchange the pseudo-range corrections to make positioning improvements in a wider field.

The CMM approach is introduced in [16]. In this solution, several neighboring vehicles exchange their pseudo-range measurements to calculate relative position between each other. Moreover, this approach calculates the position covariance and applies the road constraints of all neighboring vehicles based on map data. As the ellipse created by the covariance is the same for all neighboring vehicles using the same satellites, iterative application of all road constraints reduces the possible area on the map, where the vehicles can be located.

A distributed algorithm calculating lane-precise positions is presented in [17]. As DSRC is used for this approach, only equipped vehicles can participate, and non-automotive users are excluded from the solution. Moreover, the lane information is not derived from HD maps, but the lane width is estimated. However, a lane estimation is successfully calculated, which can be used in ADAS.

While this and many other solutions focus on lane-level positioning, i.e., lateral improvement, in [18], it is shown, how relative cooperative positioning can also be applied to improve the longitudinal GNSS error. The authors also make usage of lane-precise map information to apply road constraints. While vehicles driving on different lanes on the same road can reduce the lateral GNSS error, crossing vehicles at intersections can also improve their position in longitudinal direction.

Another approach for cooperative relative positioning is introduced in [19]. The system combines standard V2X messages transmitted via DSRC, i.e., CAM messages, with a centralized positioning algorithm. The system is installed on toll bridges on highways. With the knowledge of its precise position, the system in the toll bridge can calculate the position delta and apply it on the received positions from the CAM messages. Thereby, the toll system can precisely track vehicles and associate the correct lane in the tollgate.

An application of relative positioning in an ADAS is the cooperative active blind spot assistant (CABSA) [6]. The vehicles exchange their positioning constantly through standard CAM messages. After map matching, the relative position is calculated, and it is estimated whether other vehicles enter the blind spot area of the ego vehicle. If this is the case, CABSA warns the driver.

A statistical analysis of GNSS errors and their elimination with the calculation of the double-difference of pseudo-ranges is presented in [20]. In a simulation, different GNSS errors are modelled, including multipath effects. In addition, Doppler effect analysis of moving vehicles is performed. It is shown that relative positioning can precisely eliminate most errors, except for multipath effects.

## III. MOBILE EDGE COMPUTING ARCHITECTURE

The foundation for the data exchange of the LaaS participants is the Local Dynamic Map (LDM). This distributed service architecture provides the structure and API for the service. The overall architecture is composed of several clusters. Each cluster is coordinated by an LDM core, which controls the communication between all cluster components and the other clusters. For LaaS, for example, two clusters are used. On the one hand, the cluster for the LaaS service itself and, on the other hand, an edge cloud variant of the cluster for the participants of the service.

Both within and between the clusters, communication takes place via Message Queuing Telemetry Transport (MQTT) [21], which enables messages to be classified and exchanged on the proposition of named topics. Furthermore, wildcards are possible in topic names. Thereby, it is possible for services to subscribe and respond to all participants without having to keep a specific channel open for each participant.

LDM utilizes these possibilities by setting up a service topic structure according to defined patterns, which enables the efficient and dynamic exchange of service messages. This architecture is defined in the LDM library, which, in turn, is used by participants to implement the end applications. Due to this abstraction, the user does not have to specify topics manually.

Likewise, anonymization of data is automatically performed to preserve privacy, despite the sharing of sensitive data such as current location. The distributed architecture helps to mask the user's identification details for the service. Only a randomly generated, changing ID is assigned, which is required to send the improved position back to the user.

Service providers announce their services to their local LDM core through the LDM library, which then publishes it on a predefined service announcement topic. The LDM core then takes over the task of ensuring that all clusters are also informed about the availability of the new service. Participants can then request a list of available services from their respective LDM core using the LDM library.

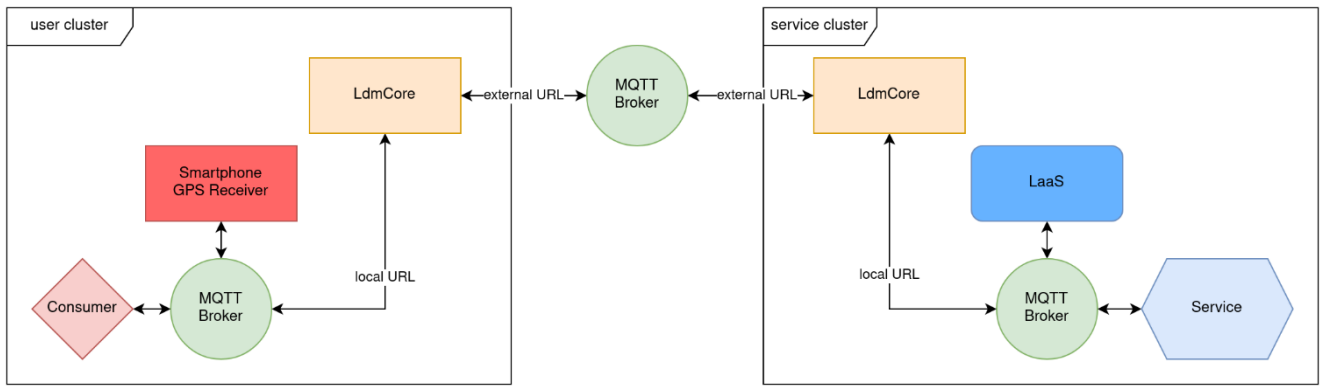

Fig. 1  LDM architecture for LaaS

Consequently, the participants receive a comprehensive description of the service from the core, as well as all the information needed to start an interaction with it. Accordingly, a participant can then prepare for the message exchange by creating the appropriate subscriptions and publishers to receive the response from the service and send the data to the services.

After completing these preparations, the participant can then send a corresponding service request to the local LDM core. This forwards the message to the respective service independently of the cluster in which it exists. The service can now also set up the publishers and subscriptions if these have not already been initiated.

From this point on, the service is ready for use by the subscriber. The participant can now send data to the service via the internally defined topics that have been adapted for the participant. The LDM library takes care of encryption. This communication also runs via the corresponding LDM core instances, so that neither the service nor the participant receives any detailed identification features of the other. The service can then be used by the participant according to the request-response principle.

In the case of LaaS, the position data of the subscriber is transmitted, and the improved position is returned by the service. Due to the wildcard topics, it is not necessary to log off explicitly from the services as a participant; it is simply sufficient to stop sending input. Furthermore, the use of the MQTT standard also makes it possible to use the Quality of Service Polices to ensure that even in error-prone networks the important messages arrive on time and meta messages, e.g., the service list, are cached efficiently and are thus immediately available to new participants.

Fig. 1 illustrates the described architecture of the LDM in interaction with MQTT from the perspective of a participant. The left box contains the user cluster. Within this cluster, all components are connected via an MQTT broker. The LDM core of the cluster links the components with other clusters. This communication between the clusters is handled by another MQTT broker. Each LDM core is connected to this broker and thus integrated into the global system. The box on the right shows the service cluster with the LaaS service. This is basically structured in the same way as the user cluster. A broker connects all components in the cluster and an LDM core takes over the coordination of the components and the communication with the external clusters.

The abstraction into clusters enables modular deployment. Many of the applications, e.g., LaaS, are time-critical; too high end-to-end latency would send back outdated data and thus provide no value to the client. By using mobile edge cloud clusters, both the entry point of the clients into the LDM system and the services can be offered in a geographically optimized manner, thus creating the physical prerequisite for low latency. The simple and easily parallelizable communication protocol enables the LDM cores to communicate efficiently with the clients and services. Nevertheless, the abstraction over several clusters also provides the necessary precautions to allow communication to take place anonymously.

## IV. Location as a Service

The GNSS location information derived from LaaS participants, i.e., a group of road users within spatial proximity, is used cooperatively to improve their positioning. This is achieved by combining two recursive state estimation techniques – a Kalman filter and a discrete Bayes filter. To obtain lane precise localization, a group of road users is mapped onto the road layout by probabilistically assigning them lanes. LaaS is fully integrated into the described LDM service architecture.

### A. GNSS Error Model

The accuracy of positions derived from GNSS systems such as GPS are subject to different error sources, often resulting in deviations larger than 5 meters.

*1) Satellite based errors* are caused by inaccuracies local to the satellites sending GNSS signals to the receiver. This type of error is typically caused by clock drift and orbit errors.

*2) Signal propagation errors* are caused by atmospheric influences on the signal while travelling from space to the receiver, e.g., humidity, atmospheric pressure, but also the reflection of the signals at surrounding structures like buildings (multipath propagation).

*3) Receiver errors* are hardware and software-based errors that occur at the receiver device.

Individually, these errors might seem small. However, the sum of the resulting errors often makes the use of raw GPS

measurements unsuitable for ADAS like Green Light Optimal Speed Advisory (GLOSA) or collision warning systems.

The measurements of LaaS participants are derived from GPS receivers, e.g., smartphones, or are simulated based on the errors described above. A GPS measurement in the LaaS architecture contains the following parameters.

- Timestamp in milliseconds
- Position (latitude, longitude, altitude)
- Heading (degrees)
- Speed (m/s)

*B. Map Information*

Implementing LaaS requires highly accurate map information, provided by so-called HD maps. In contrast to traditional map models, which describe only the road network on a node-edge model, they provide precise information about lanes. Accurate knowledge of the number, width, and connecting links of all lanes is vital for the positioning of participants, especially vulnerable road users (VRU).

That is why for the LaaS service, a highly detailed HD map was created. It contains all necessary information on the road and its lanes, as well as sidewalks and bike lanes. The map information is stored in a hierarchical structure, dividing roads into heavily linked map elements. Each element describes a piece of the road and consists of a constant number of car and bike lanes, sidewalks, and parking spaces. The information on the lanes is denoted in latitude and longitude, allowing for a precision in the centimeter range.

*C. Kalman Filter*

Apart from the receiver error and errors caused by multipath propagation, the errors of GPS receivers in proximity are assumed to be approximately the same. To mitigate the errors of individual GPS receivers, a basic Kalman filter, as described in [22] is used. It aims to smooth out the receiver noise and reduce leaps in the position measurements. Therefore, each measurement is filtered before being processed further. This way, the remaining errors are limited to the satellite and signal propagation errors, i.e., the errors that are the same for receivers within a spatially close range. The filter is tuned by approximating the sensor accuracy and its state space consists of the following parameters.

- UTM Easting (m)
- UTM Northing (m)
- X-Velocity (m/s)
- Y-Velocity (m/s)

*D. Lane Positioning*

A single GPS receiver does not provide the accuracy needed to determine on which lane it is located. Even after smoothing out the noise caused by errors in the receiver device with the Kalman filter, the satellite and signal propagation errors remain. For this reason, the positioning data of multiple road users within proximity are used cooperatively to compute the probability of a user occupying a specific lane. After filtering the GPS measurements of a group of spatially close road users, they all exhibit approximately the same position errors. Thus, they form an accurate depiction of how they are

$$\mathbf{H} = \begin{cases} h_0 = ((v_0 \to l_0, v_1 \to l_0), & p), \\ h_1 = ((v_0 \to l_1, v_1 \to l_0), & p), \\ h_2 = ((v_0 \to l_0, v_1 \to l_1), & p), \\ h_3 = ((v_0 \to l_1, v_1 \to l_1), & p) \end{cases}$$

Fig. 2  Hypotheses H consisting of all possible lane assignments

located relative to each other. For example, the distance between two cyclists can be determined accurately after filtering their position measurements, even though both positions have a large absolute error.

This property, together with the assumption that the road users are either moving on the sidewalk, on car lanes or on bike lanes, are used to evaluate the probabilities of all possible lane assignments of the road users onto the road layout.

The accurate lane location information from the HD map allows utilizing a discrete Bayes filter to derive lane probabilities for each road user within a road segment. The state space is represented by a set of hypotheses **H**. Each hypothesis $h$ describes one possible mapping of all users within the road segment, where a user $v$ is assigned to a lane $l$, as shown in Fig. 2.

The discrete Bayes algorithm continuously updates the probabilities $p$ for each set of lane assignments $h$ whenever new measurements arrive. Initially, they are evenly distributed. For each set of new measurements, the discrete Bayes algorithm computes the new probabilities in the *prediction* step, followed by the *update* step.

The prediction step updates the probabilities purely based on the Bayes filter result of the previous time step and its corresponding measurements. In this step, for all hypothesis $h$ the heading angle of each road user is compared to the direction angle of the assigned lane. A low difference between those angles indicates that the road user remains on the same lane, increasing the probability of $h$. A large difference indicates a lane change, increasing the probabilities of the hypotheses that contain the lane assignment towards which the road user is headed.

After predicting new probabilities for each hypothesis, the probabilities are updated in the update step. In this step, the most recent measurements of the road users are considered, i.e., the relative distances between them. All possible combinations of lane assignments of the users are iterated through by mapping each road user $v$ onto an assigned lane $l$. Each realized mapping is considered as hypothesis $h$. Then, for each $h$, the relative distances between the road users are computed. These distances are compared to the relative distances between the originally measured positions filtered by the Kalman filter. Depending on the size of the difference, the probability for each hypothesis is increased or decreased.

After the computation is completed, the hypothesis with the highest probability is selected. It most accurately resembles the originally measured relative distances between the road users, while placing them onto viable spots of the road layout. At last, the lane positions of the road users from the selected hypothesis are taken as the improved positions.

V. EVALUATION

The evaluation is conducted using a test scenario consisting of approximately 300 simulated GPS

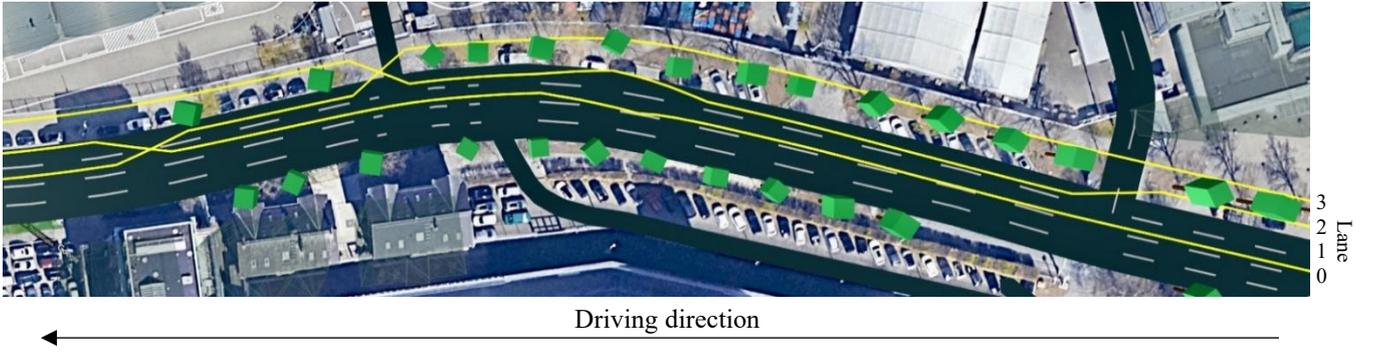

Fig. 3   Ground truth traces of the test scenario

measurements, generated with the simulation software PHABMACS [23]. Unlike real-world data, the simulator provides a ground truth. In this scenario, the movement of two Vulnerable Road Users (VRU), i.e., cyclists, and a car is emulated as they navigate a road comprised of four distinct lanes: two main vehicular lanes (referred to as lane 0 and lane 1), a parking lane (lane 2), and a bicycle lane (lane 3). The scenario covers lane transitions and includes a curve. The chosen setting for this test scenario is in the vicinity of Messe Hamburg. The ground truth paths, prior to the introduction of the GPS errors, are depicted in Fig. 3.

The car starts on lane 0 and changes to lane 1 briefly before the end of the scenario. The first cyclist starts on lane 2, switches to the main lane 1 after 20 seconds and changes to the lane 3, the cyclist lane, after another 40 seconds. The second cyclist starts on lane 3 and switches to lane 1 after 60 seconds, before moving to lane 0 towards the end of the scenario. In Fig. 3, the participating road users can be seen initially on the right side, from bottom to top, driving towards the left side during the scenario.

The simulation measures the positions, speed, and heading of the road users with virtual GPS sensors. To emulate GPS-induced errors from the receivers, multiple error models are integrated into the simulation. First, a consistent GPS offset (bias) is applied to the ground truth traces generated for each road user. This offset is defined as -8 meters east and +8 meters north, leading to an average Euclidean error of approximately 11.3 meters, which is a typical bias for GNSS [24]. This bias aims to replicate the uniform GPS discrepancies observed among receivers in close spatial proximity.

Subsequently, an error, derived from a Gaussian distribution, is introduced to each measurement point. The receiver noise for GPS is usually around 0.5 meters. Despite this, the algorithm is tested with increasing position errors, as effects close to the receivers like multipath propagation also affect the measurement errors.

The standard deviation of the distribution, represented as $\sigma_{pos}$, is configured within a range of 0 to 4 meters for both the easting and northing error. Furthermore, errors in velocity (meter per second) and in heading (radians) as Gaussian noise is added with $\sigma_{vel}$ values of 0.25 m/s in the configurations with lower position errors or 0.5 m/s in the configurations with higher position errors. The standard deviation for the heading $\sigma_{heading}$ is set fixed at 0.05°. Typically, for GPS, the errors in velocity depend on the position errors, i.e., higher position errors yield higher velocity errors. As the methods to compute velocities vary between different receiver devices, two different standard deviations for velocity are assumed. The GPS measurements are transmitted as messages in the LDM architecture to the LaaS service.

To illustrate the impact of the Kalman filter in LaaS, the implementation is tested on both unfiltered and filtered measurements. To evaluate the results of the simulation run, two metrics are used to compare the original error values with the reduced errors. The Root Mean Square Error (RMSE) quantifies the discrepancy between the ground truth and the improved positions. The average distance between the ground truth and the improved positions is calculated using the Average Euclidean Distance (AED).

Table I presents the mean results of 20 simulation runs for each specified error configuration. Lane assignments are assessed based on the ratio of correctly assigned lanes, denoted as "accuracy". Overall, the findings reveal that the LaaS lane positioning algorithm decreases the AED by over 55% across all simulated error models.

TABLE I.   RESULTS

| Kalman filter | $\sigma_{pos}$ m | $\sigma_{vel}$ m/s | RMSE$_x$ m | RMSE$_y$ m | AED m | Acc. |
|---|---|---|---|---|---|---|
| — | — | — | -8.00 | 8.00 | 11.3 | — |
| ✗ | 0.5 | 0.25 | 2.02 | 2.76 | 2.94 | 0.91 |
| ✓ | 0.5 | 0.25 | 1.93 | 2.58 | 2.75 | 0.91 |
| ✗ | 1.0 | 0.25 | 2.07 | 2.90 | 3.07 | 0.87 |
| ✓ | 1.0 | 0.25 | 1.95 | 2.70 | 2.83 | 0.88 |
| ✗ | 2.0 | 0.5 | 2.30 | 3.30 | 3.47 | 0.79 |
| ✓ | 2.0 | 0.5 | 2.11 | 2.95 | 3.00 | 0.80 |
| ✗ | 4.0 | 0.5 | 3.62 | 4.42 | 4.90 | 0.68 |
| ✓ | 4.0 | 0.5 | 3.10 | 3.70 | 4.10 | 0.66 |

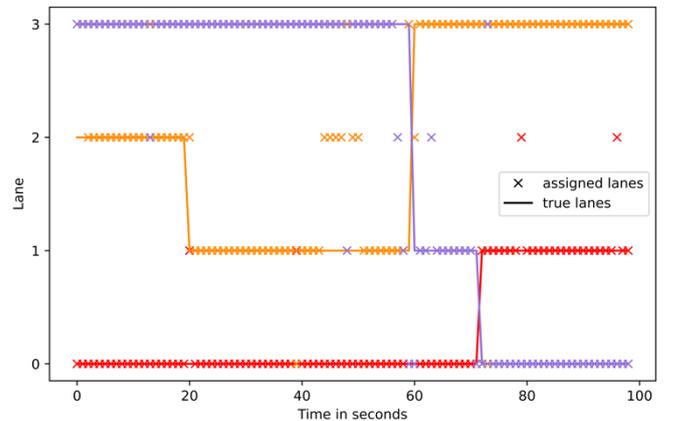

Fig. 4   Lane assignments computed by the lane positioning algorithm

Fig. 4 illustrates the lane assignments over time, as determined by the lane positioning algorithm. The simulated GPS measurements retain a bias of -8 meters east and +8 meters north, complemented by Gaussian noise parameters set at $\sigma_{pos}$ = 1.0 m, $\sigma_{vel}$ = 0.25 m/s, and $\sigma_{heading}$ = 0.05°. This figure highlights the accurate mapping of three road users to their respective true lanes.

## VI. SUMMARY

With LaaS, a system is presented that allows not only automotive vehicles but also all users with a device that has cellular connection and a GNSS receiver, e.g., VRUs, to improve their position with lane-level precision. The solution is implemented as a MEC cloud service to reduce latency and to allow time critical ADAS using the improved position.

Road users who want to participate, publish their position to the service in an anonymized way. The service collects the positions of several users and applies a lane-matching algorithm to each one. Before the lane matching, a Kalman Filter is applied to smoothen the GNSS tracks of all users.

As positions in close vicinity share the same GNSS error, the relative positions between them are highly accurate. For the calculation of the lane assignments, a Bayesian Filter is used. In each step, the position of a reference user is assigned to different lanes in an HD map, including sidewalks and cycle lanes. Consequently, the positions of all close users are assigned to different lanes. Thereby, the necessary lateral shift compared to the original relative position is measured.

The combination of lane assignments that yields the least combined lateral shift distance is assigned the highest probability. The LaaS system responds to all users with a corrected absolute position. With the usage of an HD map, a precise absolute position is derived for all users.

The evaluation in a simulation environment uses different error models. The experiments show that in all error configurations, the overall error is significantly reduced and an accuracy of more than 90% can be achieved for the lane assignment. With this accuracy, many ADAS, e.g., emergency vehicle warning, can be realized.

As navigation and assistance systems for cyclists and micro-mobility are increasingly common, real-world evaluation is planned to show the accuracy of LaaS. As no ground-truth is available for GPS receivers, this evaluation can make usage of pre-defined routes. Test-drives with real cars and cyclists show that the accuracy is very high. For further evaluation, also other error sources of GPS, e.g., urban canyons, can be taken into account. These errors can be mitigated by combining cooperative positioning with other localization techniques, e.g., LIDAR localization [18].

In a next step of implementation, LaaS can be combined with other MEC services to improve the safety and efficiency of traffic. While ADAS in general are implemented on the user side, some ADAS can also be realized as MEC services. For example, a cloud based GLOSA can calculate approach times to traffic lights for the participants.

## REFERENCES

[1] "J3016," SAE Int. J. Commer. Veh., 2012.
[2] K. Czarnecki, "Operational Design Domain for Automated Driving Systems-Taxonomy of Basic Terms," Waterloo Intell. Syst. Eng. Lab, Univ. Waterloo, Canada, pp. 1–22, 2018.
[3] T. Binder et al., "Assistenzsysteme in neuer Dimension," Sonderprojekte ATZ/MTZ, vol. 21, no. S1, 2016.
[4] D. Naudts et al., "Vehicular communication management framework: A flexible hybrid connectivity platform for CCAM services," Futur. Internet, vol. 13, no. 3, 2021.
[5] O. Sawade and I. Radusch, "Survey and classification of cooperative automated driver assistance systems," in 2015 IEEE 82nd Vehicular Technology Conference, VTC Fall 2015 - Proceedings, 2016.
[6] O. Sawade, B. Schaeufele, J. Buttgereit, and I. Radusch, "A cooperative active blind spot assistant as example for next-gen cooperative driver assistance systems (CoDAS)," in IEEE Intelligent Vehicles Symposium, Proceedings, 2014.
[7] K. Massow, I. Radusch, and R. Shorten, "A Numerical Study on Constant Spacing Policies for Starting Platoons at Oversaturated Intersections," IEEE Access, vol. 10, pp. 43766–43786, 2022.
[8] I. Llatser, T. Michalke, M. Dolgov, F. Wildschutte, and H. Fuchs, "Cooperative automated driving use cases for 5G V2X communication," in IEEE 5G World Forum, 5GWF 2019 - Conference Proceedings, 2019.
[9] B. Lehmann, H. J. Günther, and L. Wolf, "A Generic Approach towards Maneuver Coordination for Automated Vehicles," in IEEE Conference on Intelligent Transportation Systems, Proceedings, ITSC, 2018, vol. 2018-November.
[10] O. Sawade and I. Radusch, "Session-based communication over ieee 802.11p for novel complex cooperative driver assistance functions," in 21st World Congress on Intelligent Transport Systems, ITSWC 2014: Reinventing Transportation in Our Connected World, 2014.
[11] F. Bellotti et al., "TEAM Applications for Collaborative Road Mobility," IEEE Trans. Ind. Informatics, vol. 15, no. 2, 2019.
[12] L. Eiermann, O. Sawade, S. Bunk, G. Breuel, and I. Radusch, "Cooperative automated lane merge with role-based negotiation," in IEEE Intelligent Vehicles Symposium, Proceedings, 2020.
[13] N. Alam, A. T. Balaei, and A. G. Dempster, "Relative positioning enhancement in VANETs: A tight integration approach," IEEE Trans. Intell. Transp. Syst., vol. 14.1, pp. 47–55, 2012.
[14] K. Liu, H. B. Lim, E. Frazzoli, H. Ji, and V. C. S. Lee, "Improving positioning accuracy using GPS pseudorange measurements for cooperative vehicular localization," IEEE Trans. Veh. Technol., vol. 63, no. 6, pp. 2544–2556, 2014.
[15] L. Rivoirard, M. Wahl, P. Sondi, D. Gruyer, and M. Berbineau, "A Cooperative Vehicle Ego-localization Application Using V2V Communications with CBL Clustering," IEEE Intell. Veh. Symp. Proc., vol. 2018-June, no. Iv, pp. 722–727, 2018.
[16] M. Rohani, D. Gingras, and D. Gruyer, "A novel approach for improved vehicular positioning using cooperative map matching and dynamic base station DGPS concept," IEEE Trans. Intell. Transp. Syst., vol. 17, no. 1, pp. 230–239, 2016.
[17] P. Lorenz, B. Schaeufele, O. Sawade, and I. Radusch, "Recursive State Estimation for Lane Detection Using a Fusion of Cooperative and Map Based Data," IEEE Conf. Intell. Transp. Syst. Proceedings, ITSC, vol. 2015-Octob, no. 318621, pp. 2180–2185, 2015.
[18] H. Zhang, B. Schaeufele, J. N. Hark, O. Sawade, and I. Radusch, "Cooperative Longitudinal Positioning at Intersections Using DSRC," IEEE Veh. Technol. Conf., vol. 2018-Augus, 2018.
[19] M. Randriamasy, A. Cabanil, H. Chafouk, and G. Fremont, "Evaluation of methods to estimate vehicle location in Electronic Toll Collection Service with C-ITS," IEEE Intell. Veh. Symp. Proc., vol. 2018-June, no. Iv, pp. 748–753, 2018.
[20] F. De Ponte Müller, A. Steingass, and T. Strang, "Zero-Baseline Measurements for Relative Positioning in Vehicular Environments," 6th Eur. Work. GNSS Signals Signal Process., 2013.
[21] R. A. Light, "Mosquitto: server and client implementation of the MQTT protocol," J. Open Source Softw., vol. 2, no. 13, 2017.
[22] S. Thrun, B. Wolfram, and D. Fox, Probabilistic Robotics (Intelligent Robotics and Autonomous Agents). 2005.
[23] K. Massow and I. Radusch, "A Rapid Prototyping Environment for Cooperative Advanced Driver Assistance Systems," J. Adv. Transp., 2018.
[24] J. Sanz Subirana, J. M. Juan Zornoza, and M. Hernández-Pajares, GNSS Data Processing, Volume I: Fundamentals and Algorithms. ESA Communications, 2013.